\documentclass[conference]{IEEEtran}

\usepackage{cite}
\usepackage{physics}
\usepackage{hyperref}
\usepackage{lipsum} 
\usepackage{color}

\usepackage[numbers,sort&compress]{natbib} 

\usepackage{amsmath}
\usepackage{numprint} 
\usepackage{amsthm} 
\usepackage{url} 
\usepackage{textcomp} 

\usepackage{accents} 
\usepackage{amssymb}
\usepackage{amsfonts}
\usepackage{dsfont}
\usepackage[figuresright]{rotating} 
\usepackage{floatpag} 
	\rotfloatpagestyle{empty} 

\usepackage{longtable}
\usepackage{subfigure}

\usepackage{epstopdf} 
\usepackage{makeidx} 
\usepackage{upgreek} 
\makeindex 

\usepackage{times}
\usepackage{mathdots} 

\usepackage{bm}
\usepackage{bbm}

\theoremstyle{plain}

\def\bb0{{\mathbb{0}}}

\def\bb{{\boldsymbol{b}}}

\def\bh{{\boldsymbol{h}}}

\def\br{{\boldsymbol{r}}}
\def\bs{{\boldsymbol{s}}}

\def\bu{{\boldsymbol{u}}}

\def\bz{{\boldsymbol{z}}}
\def\b0{{\boldsymbol{0}}}

\def\b{{\mathrm{b}}}

\def\r0{{\mathbf{0}}}

\def\cW{\mathcal{W}}
\def\cX{\mathcal{X}}

\def\sfR{\mathsf{R}}

\def\bTheta{\bm \Theta}

\def\sfi{{\mathsf{i}}}

\def\sfn{{\mathsf{n}}}

\def\sfr{{\mathsf{r}}}
\def\sfs{{\mathsf{s}}}
\def\sft{{\mathsf{t}}}

\def\bsf0{{\bm{\mathsf{0}}}}

\def\c{{\rm c}} 

\def\N0{{N_{\mathrm{0}}}}
\def\sinc{\mathrm{sinc}}

\def\j{\mathrm{j}}

\def\fc{f_{\mathrm{c}}}

\def\lambdac{\lambda_{\mathrm{c}}}

\def\bsf{{\boldsymbol{s}_\mathrm{f}}}

\newcommand{\Trans}{\rm \scriptscriptstyle T}

\def\diag   {\mbox{\rm diag}}

\def\RIS{{\sf RIS}}

\def\I{\mathbbm{1}}

\graphicspath{{./images/}}

\begin{document}
\title{RIS in Indoor Environments: Benchmarking Against Ambient Propagation}

\author{
\IEEEauthorblockN{Masoud Sadeghian, Andrea Pizzo, Angel Lozano}
\IEEEauthorblockA{Universitat Pompeu Fabra, 08018 Barcelona, Spain}
}

\maketitle

\begin{abstract}
The improvements in received signal power brought about by a reflective intelligent surface (RIS) might be overstated if background propagation mechanisms such as reflections, scattering, and diffraction are ignored. This paper addresses this issue for non-line-of-sight indoor settings, contrasting the energy conveyed by an RIS with the energy already reaching the receiver through environmental reflections. And, to prevent artifacts, such naturally occurring reflections are not modeled via approximate methods, but rather through a rigorous physics-based formulation. It is found that the environment contributes a level of energy commensurate with that of an ideal RIS of considerable size; to have substantial impact, an actual RIS would have to generously exceed this size. 
\end{abstract}


\section{Introduction}

The possibility of controlling radio propagation environments is fuelling the interest in reflective intelligent surfaces (RISs) \cite{BasarSurvey,ZhangSurvey}. 
In particular, 
it is envisioned that a RIS can circumvent the obstruction of the line-of-sight (LOS) path and make communication possible where it otherwise would not be. 
%
However, most evaluations of the RIS benefits in this respect, and in the sense of broadly improving coverage, disregard naturally occurring signal reflections that can play a similar role. This creates the risk of overstating the advantages of deploying a RIS. Indeed, it has been shown that, for an
around-the-corner outdoor setting, reflections from building and street poles contribute as much power as a RIS of $0.3 \, \text{m} \times 0.3 \, \text{m}$ \cite{Valenzuela2022}, meaning that only a RIS considerable larger than this would have a substantial impact.

In indoor settings, ambient propagation might be of even greater importance, yet only seldom it is taken into account \cite{zhang2021fundamental,Sciancalepore2022}.
Motivated by this concern, this paper seeks to reference the power conveyed by a RIS to the power naturally reaching a receiver indoors.
To prevent artifacts, approximate methods are avoided 
and we instead resort to an exact assessment on a setting that is simple yet representative, in fact the chief building block of most indoor environments: a room. This suffices to capture the main indoor propagation mechanism at high frequencies, which is the reflection on flat surfaces.
Importantly, the LOS path is considered obstructed, as that is when both ambient propagation and a RIS become relevant.

\begin{figure}[t] 
\centering
\includegraphics[width=.85\linewidth]{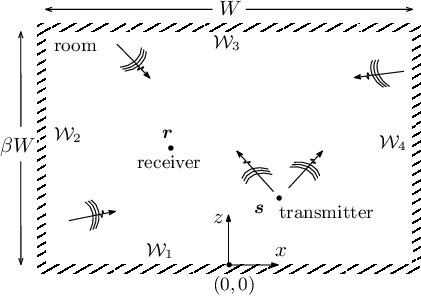}
\caption{Rectangular room with longer dimension $W$ and aspect ratio $1/\beta$, with exemplary transmitter and receiver locations. The coordinate system is also shown. 
Although not made explicit, the LOS path is blocked.} 
\label{fig:model1}
\end{figure}

\begin{figure}
\centering
\includegraphics[width=.85\linewidth]{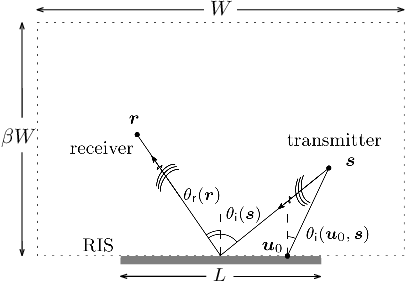}
\caption{RIS of size $L \le W$ in free space. Although not made explicit, the LOS path is blocked.}
\label{fig:model2}
\end{figure}

The paper begins by establishing the power received in an empty room when transmitter and receiver are randomly positioned (Fig. \ref{fig:model1}), then it repeats the exercise with the room replaced by a RIS (Fig. \ref{fig:model2}), and finally it draws a relationship between the two.
As a by-product of the formulation, the near-field impulse response of a RIS-assisted channel is derived and might be of independent interest.
This derivation parallels the one in \cite{9433568}, generalizing existing results where the link between transmitter and RIS and/or the link between RIS and receiver is in the far field \cite{Emil2020,9941256}.

\section{Indoor Ambient Propagation}
\label{Mirotic}

Transmission is considered at carrier frequency $\fc$, initially inside a rectangular empty room, the electromagnetic properties of whose walls are summarized by a complex refractive index
\begin{equation} \label{refractive_index}
\sfn = \sqrt{\epsilon_{\sf r} \mu_{\sf r}} ,
\end{equation} 
where $\epsilon_{\sf r}$ is the permittivity and $\mu_{\sf r}$ the permeability, both relative to free space.
The bandwidth is small enough 
that $\sfn$ can be regarded as constant thereover.
A single polarization is considered, with no environmental depolarization such that all electromagnetic quantities can be expressed as scalars \cite{tulino2003capacity}.

A 2D plane corresponding to a horizontal slice of the room is studied, given the space invariance of \eqref{refractive_index}. 
The room boundaries can then be modeled as a rectangle $ \cW = \cup_{i=1}^4  \cW_i$ where $\cW_i$ is the $i$th wall (see Fig.~\ref{fig:model1}). There are no internal reflections within each wall.
The formulation seeks to broadly encompass any situation in which the room is much larger than the wavelength, and the interest is in the relative signal powers conveyed by the environment and the RIS, rather than their absolute value.
The room's aspect ratio is $1/\beta \geq 1$, with $W$ the longer room dimension. As advanced, the LOS path is taken to be obstructed. For the sake of specificity, the results are produced for $\fc=28$~GHz, which is a particularly suitable frequency for indoor communication, yet these same results are valid for any other frequency with a properly scaled $W$.

\subsection{Image Theorem}

For an unbounded smooth wall of perfectly conducting material, the boundary conditions amount to the image theorem, which states that the reflection produced by the surface can be exactly mimicked by a mirror image of the transmitter in relation to such wall. 
The image theory readily generalizes to the intersection of walls by imaging every image in turn, until the arising images overlap  \cite[Ch.~7.4]{BalanisBookEM}.
In the case of Fig.~\ref{fig:model1}, each step $m \geq 1$ of the image-forming process adds $4m$ 
images, eventually producing a 2D lattice of images \cite{Dienstfrey2001}. 
An $m$th stage image models the propagation from transmitter to receiver via $m$ successive reflections.
Leveraging the linearity of wave propagation, the channel impulse response at $\br = (r_x,r_y)$ due to an impulsive current (point source) at $\bs = (s_x,s_z)$ is obtainable by considering each wall separately and adding all image contributions into
\begin{align} \label{image_process}
h(\br,\bs) = \sum_{m=1}^\infty \sum_{i=1}^{4m}  h_{m,i}(\br,\bs)
\end{align}
with $h_{m,i}(\br,\bs)$ the contribution of the $m$th order image in relation to the $i$th wall.
For instance, for the first-order image in relation to $\cW_1$,
\begin{align}  \label{convolution_pec_2}
h_{1,1}(\br,\bs) & = - G(\br,s_x,-s_z),
\end{align}
with ${0 < s_z <\beta W}$ and ${-W/2 <s_x < W/2}$; the receiver is similarly confined.
The sign reflects the phase inversion and
\begin{equation} \label{Green_fun}
G(\br,\bs) = \frac{\j}{4} H_0^{(1)}\!\left(\frac{2\pi}{\lambdac} \|\br-\bs\|\right)
\end{equation}
is the 2D Green's function describing a cylindrical wave, with $H_\nu^{(1)}(\cdot)$ the $\nu$th order Hankel function of the first kind, $\lambdac = \c/\fc$ the carrier wavelength, and $\c$ the speed of light.
Alternatively, \eqref{convolution_pec_2} can be rewritten as
\begin{align}  \label{convolution_pec}
h_{1,1}(\br,\bs) & = \iint_{-\infty}^{\infty} j_{1,1}(\bu,\bs) \, G(\br,\bu) \, d\bu
\end{align}
where 
\begin{align} \label{image_source_pec}
j_{1,1}(\bu,\bs) = - \delta(u_z + s_z) \delta(u_x-s_x)
\end{align}
is the current density of the image source, for $\bu = (u_x,u_z)$.
The delta functions confine this current to the plane at $-s_z$, parallel to $\cW_1$, shifting horizontally in correspondence to $s_x$.

\subsection{Generalized Image Theorem}


The image of a source appears as sharp as the original only under perfect reflection.
With walls made of non-perfectly-conducting materials,
 the image is blurred by a convolution with the impulse response of the material \cite{PizzoTWC22}. 
For 
linear transverse electric polarization (electric field parallel to the wall), the image current specializes to the inverse Fourier transform
\begin{equation} \label{material_response}
R(u_x) = \frac{1}{2 \pi} \int_{-\infty}^{\infty} \sfR(k_x)  \, e^{\j k_x u_x} \, dk_x
\end{equation}
of the wavenumber spectrum 
\begin{align} \label{Fresnel_spectrum}
\sfR(k_x) 
= \frac{\mu_{\sf r} \sqrt{1-\left(\frac{\lambdac}{2\pi} k_x\right)^2} - \sqrt{\sfn^2-\left(\frac{\lambdac}{2\pi} k_x\right)^2}}{\mu_{\sf r} \sqrt{1-\left(\frac{\lambdac}{2\pi} k_x\right)^2} + \sqrt{\sfn^2-\left(\frac{\lambdac}{2\pi} k_x\right)^2}} 
\end{align}
where $|k_x| \le {2\pi}/{\lambdac}$ once evanescent waves are ruled out.
Its magnitude is less than unity, consistent with conservation of energy.
A perfectly conducting material arises as a special case of \eqref{Fresnel_spectrum} when $\sfn \to \infty$; then, $\sfR(k_x) = -1$ and $R(u_x) = -\delta(u_x)$.


The expression of the first-order image in relation to $\cW_1$, given in (\ref{image_source_pec}) for a perfect conductor, generalizes to \cite[Eq.~23]{PizzoTWC22}
\begin{align} \label{image_source}
j_{1,1}(\bu,\bs) = \delta(u_z +s_z) R(u_x-s_x)
\end{align}
for every $\bu$.
The channel response for this reflection is then given by the spatial convolution
\begin{align}  \label{convolution}
h_{1,1}(\br,\bs)  & = \int_{-\infty}^{\infty} R(u_x-s_x) \, G(\br,u_x,-s_z) \, du_x ,
\end{align}
which generalizes \eqref{convolution_pec_2} to arbitrary materials.

Building on this system-theoretical description of the reflection phenomena, the overall wavenumber response 
at step $m$ emerges as the product of the individual responses at steps $1$ through $m$, namely $[\sfR(k_x)]^m$. 
The conjunction of imperfect reflectivity by materials of interest and the longer distances travelled by the signals generated by higher-order images allows truncating \eqref{image_process} to a finite order $M$. 
The channel power gain is then
\begin{align} \label{image_process_power}
		|h(\br,\bs)|^2 \approx \left| \sum_{m=1}^M \sum_{i=1}^{4m}  h_{m,i}(\br,\bs)\right|^2.
\end{align}
To gauge the impact of the truncation, Fig.~\ref{fig:Combined_PowerSignal} depicts the cumulative distribution function (CDF), over the locations of transmitter and receiver, of $|h|^2$. The walls are made of concrete, whose refractive index is $\sfn=5.31-\j0.3106$ at $\fc=28$~GHz \cite{itu_Matlab}.
Truncation at $M=3$ ensures a rather precise characterization, and even $M=2$ is rather satisfactory.


\subsection{Phase Randomization}

\begin{figure}[t] 
	\centering
	\includegraphics[width=.999\linewidth]{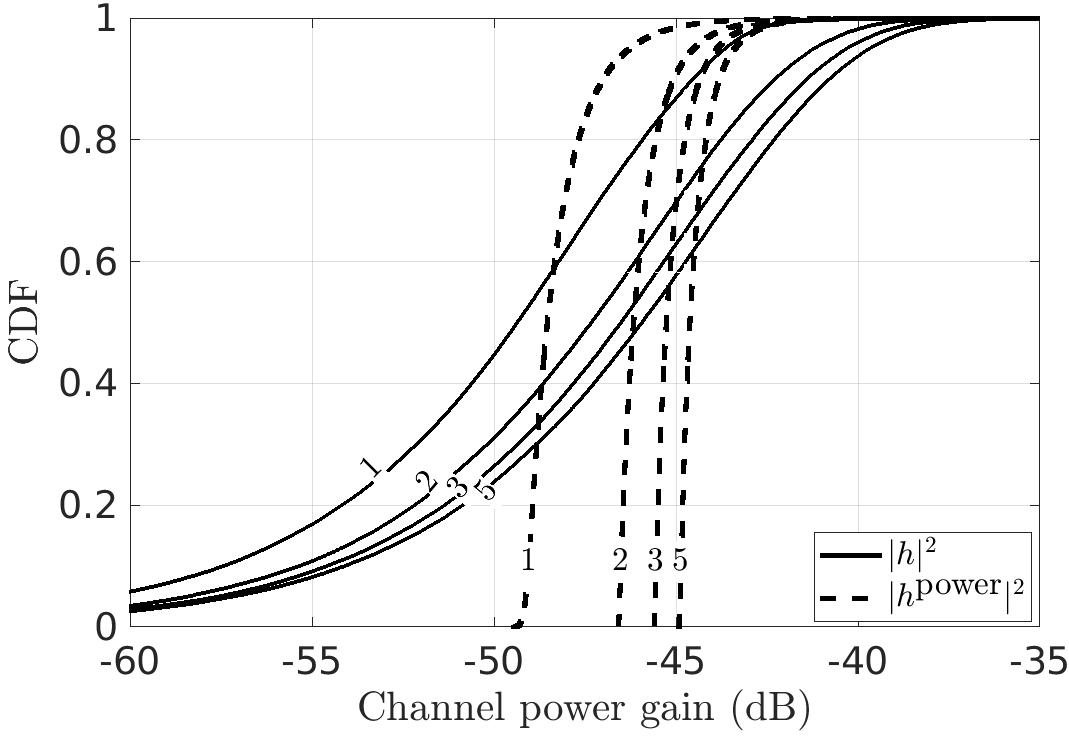}
	\caption{CDF of $|h|^2$ and $|h^{\sf \scriptscriptstyle power}|^2$ for $W=10$ m, $\beta=1$, and $M=1,2,3$ and $5$, with walls made of concrete at $\fc=28$~GHz. Scaling $W$ would alter the power gain in inverse proportion.}
	\label{fig:Combined_PowerSignal}
\end{figure}

Each of the terms being added in (\ref{image_process}) has a phase that depends, besides the refractive index and the angles of incidence onto the walls, on the total path length relative to the wavelength. The resulting impulse response is highly selective in space and frequency, which is problematic for two reasons:
\begin{itemize}
\item This selectivity is rather sensitive and could be outright deceptive, as external factors (say thermal expansion) could alter the path lengths by a nonnegligible share of the wavelength \cite{Sciancalepore2022}.
\item With strong channel coding, the performance is in essence averaged over the signal bandwidth and, if transmitter and/or receiver are in motion, further over time \cite{lozano2012yesterday}.
\end{itemize}
A representation where these two issues are skirted is one where the addition, rather than over the complex terms as in (\ref{image_process}), is over their powers \cite{rappaport2010wireless}; this gives an alternative impulse response $h^{\sf \scriptscriptstyle power}$  satisfying 
\begin{equation}
\label{Gavi}
|h^{\sf \scriptscriptstyle power}(\br,\bs)|^2 \approx \sum_{m=1}^M \sum_{i=1}^{4m} |h_{m,i}(\br,\bs)|^2 ,
 \end{equation}
 where the approximation is again with respect to the truncation at $M$.
As far as the average power goes, the above can be interpreted as applying a (uniform) phase randomization to the terms prior to their combining.
The CDF of $|h^{\sf \scriptscriptstyle power}(\br,\bs)|^2$ is also shown in Fig.~\ref{fig:Combined_PowerSignal}. Both the upper and especially the lower tails are subdued relative to $|h|^2$, as the destructive and constructive combinations of terms therein give way to a mere addition of powers. This confirms that $h^{\sf \scriptscriptstyle power}$ is a good proxy for the performance with wideband signals.

With a view to a characterization that is as general as possible, we resort to the extremes represented by $h$ and $h^{\sf \scriptscriptstyle power}$ to bracket the performance of the indoor benchmark.

 
%

\section{RIS-Aided Propagation}


Let us now remove the room and introduce a perfectly conducting RIS in its stead.
Again, the LOS path is blocked, but otherwise the transmission is in free space.
Without loss of generality, the RIS can be aligned with the $x$-axis (see Fig.~\ref{fig:model2}) 
and its size is 
$L \le W$.
Diffraction at the endpoints of the RIS is disregarded, as its effect is negligible when the RIS is electrically large.

First, the impulse response of the reflection off a bounded 
perfectly conducting surface is derived. Then, the result is extended to a RIS of the same size and material, making the impulse response controllable. 

\subsection{Perfectly Conducting Bounded Surface}

The boundary conditions in Fig.~\ref{fig:model2} cannot be replicated through the image theorem, 
and alternative methods are required to characterize the reflection. 
The standard physics approach to this problem is
Huygen's principle \cite{Pfeiffer2013,Decker2015}, whereby a field impinging on a surface induces a current thereupon that acts as a secondary source. 
This source fictitiously reproduces the radiation of the actual source by satisfying the same boundary conditions over the reflecting surface \cite[Ch.~7.10]{BalanisBookEM}.
For instance, the current density induced at every $\bu$  by a point source at $\bs$ is  (see App.~\ref{app:PHY})
\begin{equation} \label{surface_current_point}
j_L(\bu,\bs) = \frac{2}{\mu_0} \delta(u_z) \, \I_{\left[\!-\frac{L}{2}, \frac{L}{2}\!\right]}(u_x)  \, \frac{\partial}{\partial u_z} G(\bu,\bs) 
\end{equation}
where $\mu_0 = 4\pi \cdot 10^{-7}$ is the permeability of free space,
$\I_\cX(\cdot)$ is the indicator function of a set $\cX$, and
\begin{align} \label{par_der_Green}
\frac{\partial}{\partial u_z} G(\bu,\bs) & = \frac{\j \pi}{2\lambdac} \frac{|u_z - s_z|}{\|\bu-\bs\|} H_1^{(1)}\!\!\left(\frac{2\pi}{\lambdac} \|\bu-\bs\|\right) ,
\end{align}
which is derivable from \eqref{Green_fun} by using the chain rule and invoking  $\frac{\partial}{\partial z} H_0^{(1)}\!(z) =  -H_{1}^{(1)}\!(z)$ \cite[Eq.~1.2.34]{ChewBook}.
Evaluating \eqref{par_der_Green} at $u_z=0$ yields,
\begin{align} \label{par_der_Green_u0}
\frac{\partial}{\partial u_z} G(\bu_0,\bs) & = \frac{\j \pi}{2\lambdac} \frac{|s_z|}{\|\bu_0-\bs\|} H_1^{(1)}\!\!\left(\frac{2\pi}{\lambdac} \|\bu_0-\bs\|\right)
\end{align}
with $\bu_0 = (u_x,0)$. 

The delta function in \eqref{surface_current_point} confines the induced current density to the plane of the reflecting surface while the indicator function windows that current to the space occupied by the surface. In turn, 
\begin{equation} \label{cosine}
\frac{|s_z|}{\|\bu_0-\bs\|} = \cos \theta_\sfi(\bu_0,\bs) \ge 0
\end{equation}
is the cosine of $\theta_\sfi(\bu_0,\bs) = \angle(\hat{\bz},\bu_0-\bs) \in [0,\pi/2]$, shown in Fig.~\ref{fig:model2} between the surface normal $\hat{\bz}$ and the vector connecting each secondary point source at $\bu_0$ to the actual point source at $\bs$.  
This cosine yields one when the propagation is focused about the $z$-axis, while it approaches zero for very shallow angles, corresponding to the relative projection between transmitter and reflecting surface \cite{HeedongIRS}.
Finally, the term $2\frac{\partial}{\partial u_z} G(\bu_0,\bs)$ in \eqref{surface_current_point} is directly connected to the Neumann boundary conditions for a perfect conductor \cite{PlaneWaveBook} (see also App.~\ref{app:PHY}).

\subsection{Near- and Far-Field Impulse Responses}

The response at $\br$ elicited from the reflecting surface due to an impulsive current at $\bs$ is the spatial convolution
\begin{align} 
h_L(\br,\bs) & = \iint_{-\infty}^\infty  G(\br,\bu) j_L(\bu,\bs) \,  d\bu \\ \label{channel_surface}
& = \frac{2}{\mu_0} \int_{-L/2}^{L/2}  G(\br,\bu_0) \frac{\partial}{\partial u_z} G(\bu_0,\bs)  \,  du_x
\end{align}
where $j_L(\bu,\bs)$ was replaced by its expression in \eqref{surface_current_point}, and the delta function was removed by integrating over $u_z$.
Recalling 
\eqref{Green_fun} and \eqref{par_der_Green_u0}, 
the near-field response emerges as
\begin{align} \notag
h_L(\br,\bs) & = - \frac{\pi}{4\mu_0\lambdac} \int_{-L/2}^{L/2}  H_0^{(1)}\!\!\left(\frac{2\pi}{\lambdac} \|\br-\bu_0\|\right) \cos\theta_\sfi(\bu_0,\bs)  \\
& \hspace{2cm} \label{near_field}  
 \cdot H_1^{(1)}\!\!\left(\frac{2\pi}{\lambdac} \|\bu_0-\bs\|\right) \,  du_x.
\end{align}
For any fixed $u_x$, the channel response is a cascade of two Hankel's functions modeling the LOS propagation of cylindrical waves: the link between the transmitter and every secondary source on the reflector, and the link between every secondary source and the receiver. The total response is readily obtained by superposition. 

The far-field response of the channel can be obtained from \eqref{channel_surface} for a reflecting surface sufficiently far from transmitter and receiver. The connection between the two representations, expounded in App.~\ref{app:FF}, leads to the far-field response
\begin{align} \label{far_field} 
h_L(\br,\bs) & \approx E_\sfi(\bs) \cos\theta_\sfi(\bs) \, \frac{e^{\j \frac{2\pi}{\lambdac} \|\br\|}}{\sqrt{2\pi \|\br\|}} \\& \hspace{1cm} \notag
\frac{L}{\lambdac} \, \sinc\!\left(\frac{L}{\lambdac} \big(\sin\theta_\sfi(\bs) - \sin\theta_\sfr(\br)\big)\right)
\end{align}
where 
$\theta_\sfi(\bs) = \angle(\hat{\bz},\bs)$ and $\theta_\sfr(\br) = \angle(\hat{\bz},\br)$ are, respectively, the incident angle between $\hat{\bz}$ and $\bs$, and the reflected angle between $\hat{\bz}$ and $\br$ (see Fig.~\ref{fig:model2}). In turn,
\begin{equation}
E_\sfi(\bs) = \frac{\lambdac}{2 \mu_0} \frac{e^{\j \frac{2\pi}{\lambdac} \|\bs\|}}{\sqrt{2\pi \|\bs\|}}
\end{equation}
is the amplitude of the incident field, whose power decays with $2\pi\|\bs\|$ as per the cylindrical wave expansion.

Asymptotically in the electrical aperture of the reflecting surface, using $\lim_{a\to\infty} a \, \sinc(a x) = \delta(x)$,
\begin{equation}
\lim_{L/\lambdac\to\infty} h_L(\br,\bs) = h_\infty(\br,\bs)
\end{equation}
with
\begin{equation}
h_\infty(\br,\bs) = \frac{\lambdac}{2 \mu_0} \frac{e^{\j \frac{2\pi}{\lambdac} \|\bs\|}}{\sqrt{2\pi \|\bs\|}}
\frac{e^{\j \frac{2\pi}{\lambdac} \|\br\|}}{\sqrt{2\pi \|\br\|}} \cos\theta_\sfi(\bs) 
\delta\big(\theta_\sfi(\bs) - \theta_\sfr(\br)\big).
\end{equation}
This is nonzero only for $\theta_\sfr(\br)=\theta_\sfi(\bs)$, in agreement with Snell's law for an unbounded reflector made of a perfectly conductive material. 
For finite reflector apertures, \eqref{far_field} accounts for all the spurious reflections at nonspecular angles.


From \eqref{far_field}, the channel power gain for far-field observations equals
\begin{align} \notag
|h_L(\br,\bs)|^2 & \approx \frac{E_\sfi^2(\bs)}{2\pi\|\br\|} \cos^2\theta_\sfi(\bs) \left(\frac{L}{\lambdac}\right)^{\!2} \\&\hspace{1cm} \cdot \sinc^2\!\left(\frac{L}{\lambdac} \big(\sin\theta_\sfi(\bs) - \sin\theta_\sfr(\br)\big)\!\!\right) ,
\end{align}
which equals \cite[Eq.~4]{Emil2020}.
We hasten to emphasize that 
our derivation, and the result in (\ref{near_field}), do not rely on far-field approximations for the transmitter-RIS and/or RIS-receiver links.  Also, the derivation 
would naturally adapt to 3D environments due to the generality of the Green's function formulation in \eqref{channel_surface}: cylindrical waves could be replaced by spherical ones, whose power decays with $4\pi\|\bu\|^2$ rather than $2\pi\|\bu\|$.

\subsection{RIS}

The currents induced on a regular surface and a RIS are of a different nature. The former, $j_L(\bu)$, measures the share of incident field that is reflected by the surface; it is determined by the transmitter and the material composition, over which no control is exerted.
The latter, $j_L^{\scriptscriptstyle \RIS}(\bu)$, is in principle an arbitrary function (subject, in practice, to realizability constraints). 

For a RIS made of passive elements, the performed operation can be regarded as the pointwise multiplication
\begin{equation}
j_L^{\scriptscriptstyle \RIS}(\bu) = j_L(\bu) \, e^{\j \vartheta(\bu)}
\end{equation}
and the ensuing impulse response  is obtainable in the same vein as \eqref{near_field}, namely
\begin{align} \notag
\!\! h_L(\br,\bs) & = \frac{- \pi}{4\mu_0\lambdac} \int_{-L/2}^{L/2} \! H_0^{(1)}\!\!\left(\frac{2\pi}{\lambdac} \|\br-\bu_0\|\right) \cos\theta_\sfi(\bu_0,\bs)  \\
& \hspace{1.5cm} \label{near_field_RIS}  
 \cdot H_1^{(1)}\!\!\left(\frac{2\pi}{\lambdac} \|\bu_0-\bs\|\right) e^{\j \vartheta(u_x)} \,  du_x,
\end{align}
where, recall, $\bu_0 = (u_x,0)$.
Here, the phase shifts applied at every point on the RIS, $\vartheta(u_x)$, are what renders the environment controllable from an electromagnetic perspective. 

For an $N$-element RIS, the integral representation of the impulse response in \eqref{near_field_RIS} discretizes into 
\begin{align} \label{near_field_RIS_discrete}  
h_L(\br,\bs) & = - \frac{\pi}{4\mu_0\lambdac} \frac{L}{N} \sum_{n=1}^N H_0^{(1)}\!\!\left(\frac{2\pi}{\lambdac} \|\br-\bu_{0,n}\|\right)  \\ \notag
& \hspace{0.5cm}  
 \quad \cdot \cos\theta_\sfi(\bu_{0,n},\bs)  \, H_1^{(1)}\!\!\left(\frac{2\pi}{\lambdac} \|\bu_{0,n}-\bs\|\right) e^{\j \vartheta_n}
\end{align}
where $\vartheta_n = \vartheta(n L/N)$ is obtained by sampling $\vartheta(u_x)$ uniformly over the RIS support.
A vectorization of \eqref{near_field_RIS_discrete} yields
\begin{align}  \label{near_field_RIS_discrete_vector}
h_L(\br,\bs)  =  \bh_{\sfs\sfr,L}(\br)^{\Trans} \, \bTheta \, \bh_{\sft\sfs,L}(\bs) 
\end{align}
where $\bTheta = \diag(e^{\j \vartheta_1}, \ldots, e^{\j \vartheta_N})$ contains the phase-shift coefficients  and
\begin{align}
[\bh_{\sft\sfs,L}]_n(\bs) & =  - \frac{\pi}{4\mu_0\lambdac} \cos\theta_\sfi(\bu_{0,n},\bs) \, H_1^{(1)}\!\!\left(\frac{2\pi}{\lambdac} \|\bu_{0,n}-\bs\|\right) \nonumber \\
[\bh_{\sfs\sfr,L}]_n(\br) & = \frac{L}{N} H_0^{(1)}\!\!\left(\frac{2\pi}{\lambdac} \|\br-\bu_{0,n}\|\right)
\end{align}     
are the channel between the transmitter and RIS, and the RIS and receiver, for every $\br$, $\bs$, and $n=1, \ldots, N$.
With proper channel state information, the composite channels in \eqref{near_field_RIS_discrete_vector} can be cophased such that 
\begin{align}  \label{near_field_RIS_optimal}
h_L^\star(\br,\bs)  =  \|\bh_{\sfs\sfr,L}(\br)\| \, \|\bh_{\sft\sfs,L}(\bs)\| ,
\end{align}
which, from the  Cauchy-Schwarz inequality, is indeed attainable with properly optimized phase shifts $\{\vartheta_n^\star\}$.

\begin{figure}[t] 
	\centering
	\includegraphics[width=.98\linewidth]{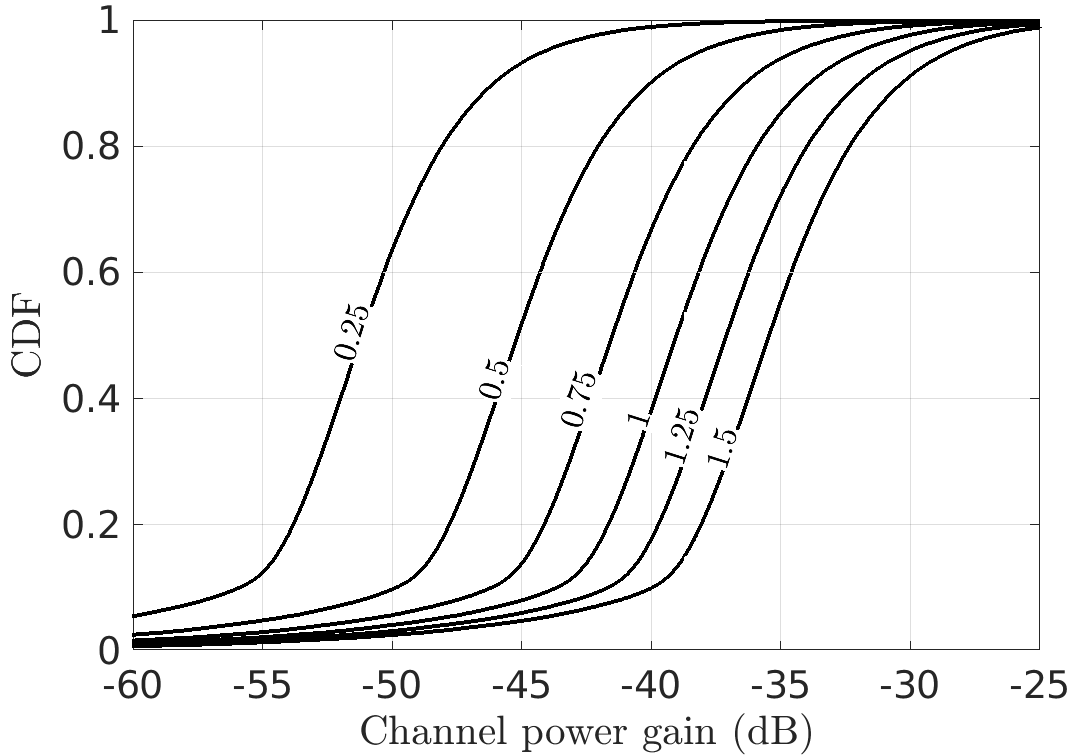}
	\caption{CDF of $|h_L^\star|^2$ for $W=10$ m, $\beta=1$, $\fc=28$ GHz, parameterized by $L =0.25,0.5,0.75,1,1.25$ and $1.5$ m, with walls made of concrete. Scaling $L$ and the space on which transmitter and receiver are positioned by a common factor would not alter the power gain.}
	\label{fig:RIS_channel}
\end{figure}

The CDF of $|h_L^\star|^2$ is depicted in Fig. \ref{fig:RIS_channel}, for various RIS sizes. The lower tail corresponds to locations from which the RIS is viewed at very shallow angles, performing poorly.

\section{RIS vs Ambient Propagation}

\begin{figure}[t] 
	\centering
	\includegraphics[width=.999\linewidth]{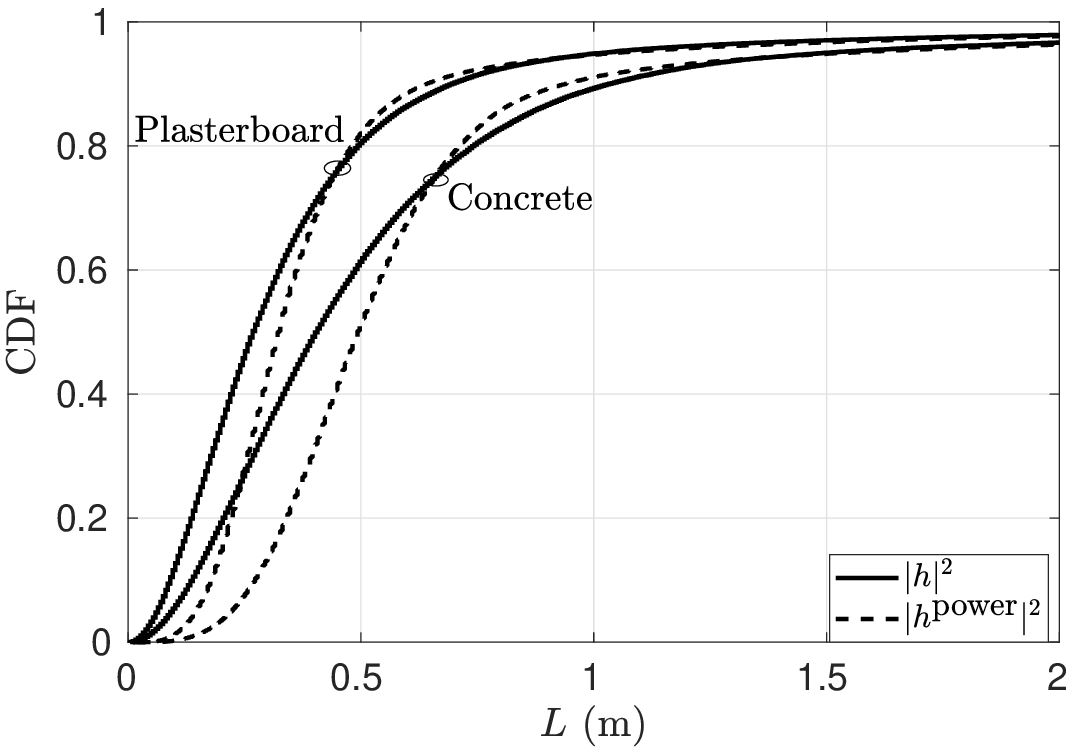}
	\caption{CDF of the RIS size delivering the same power as the ambient propagation for $W=10$ m and $\beta=1$, with either concrete or plasterboard materials, at $\fc=28$ GHz.}
	\label{fig:Comparison_W10}
\end{figure}

\begin{figure}[t] 
	\centering
	\includegraphics[width=.999\linewidth]{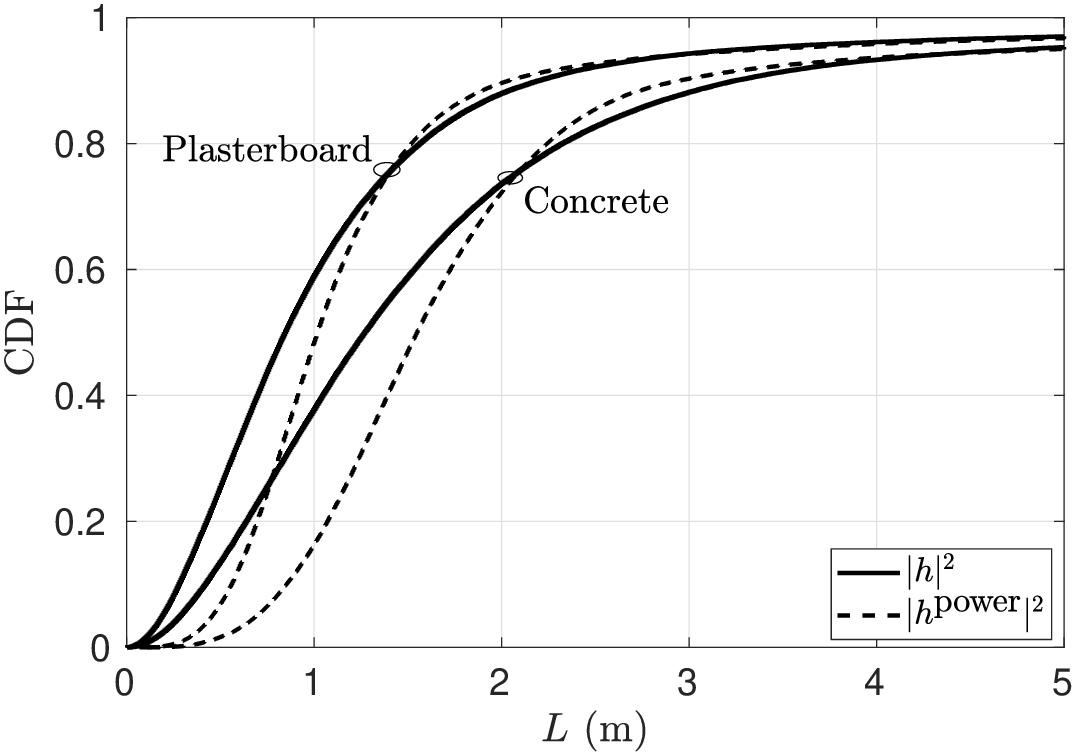}
	\caption{CDF of the RIS size delivering the same power as the ambient propagation for $W=100$ m and $\beta=1$, with either concrete or plasterboard materials, at $\fc=28$ GHz.}
	\label{fig:Comparison_W100}
\end{figure}

Armed with the formulation in the preceding two sections, the power received via a RIS can be gauged against the power received
by virtue of only ambient propagation, in both cases with the LOS path blocked.
The way we choose to articulate this comparison is by quantifying the RIS size that renders both powers equal.
Recalling the two representations of the indoor propagation derived in Sec. \ref{Mirotic}, this amounts to the value of $L$ that makes $|h|^2=|h^\star_L|^2$ or $|h^{\sf \scriptscriptstyle power}|^2=|h^\star_L|^2$.
The CDF of such $L$, taken over the position of transmitter and receiver, is presented in Figs. \ref{fig:Comparison_W10}--\ref{fig:Comparison_W100} for two distinct materials.

Accepting that the RIS may fall short of providing as much power as the environment in $10\%$ of locations (those with an overly shallow angle of incidence), for an office of $W=10$~m it is found from Fig. \ref{fig:Comparison_W10} that a RIS of size $L \approx 1$ m would be required.
For a larger indoor space, say a mall or an airport where $W=100$ m, Fig. \ref{fig:Comparison_W100} indicates that the required size swells to $L \approx 3$ m. To have a pronounced impact in the face of ambient propagation, the RIS would have to generously exceed these respective sizes in each of the environments.

The exact values change with the room's aspect ratio and the RIS position, but the qualitative observations are upheld,
and two conclusions are prompted:
\begin{itemize}
\item Very considerable RIS sizes are required to really transform the radio conditions within an indoor environment. At mmWave frequencies, this maps to truly massive numbers of controllable elements.
\item Ignoring ambient propagation is ill-advised, and RIS performance evaluations not accounting for it should be regarded with suspicion.
\end{itemize}

The characterization of the ambient propagation underpinning these conclusions is certainly based on the premises of smooth walls and of a room devoid of objects. In realistic situations, power may be scattered \cite{degli2022reradiation}, lessening the amount that reaches the receiver.
At the same time, the cophasing effected by the RIS is sure to be imperfect when based on channel estimates rather than the true channel, and with necessarily coarse phase shift resolution. Furthermore, the RIS is sure to be unable to reflect all of its impinging power. For any reasonable balance between these various effects, the above conclusions should not be compromised, and may even be reinforced.



\section*{Acknowledgment}

Work supported by the European Union-NextGenerationEU, by ICREA, by AGAUR SGR 00469,
and by the Horizon 2020 MSCA-ITN-METAWIRELESS Grant Agreement 956256.

\appendices

\section{} \label{app:PHY}

The electromagnetic physical equivalent \cite[Ch.~7.10]{BalanisBookEM} is specialized next to linearly polarized transverse electric fields, which are amenable to a scalar description \cite[Ch.~2]{ChewBook}.

\subsection{Boundary Conditions for Transverse Electric Polarization}

The wave equation driven by a current density $j(\bu)$, $\bu = (u_x,u_z)$, in an isotropic medium of space-dependent permeability $\mu(\bu)$ and permittivity $\epsilon(\bu)$ is  \cite{ChewBook}
\begin{equation} \label{wave_eq}
 \div \frac{1}{\mu(\bu)} \grad e(\bu) + \left(\frac{2\pi \c}{\lambdac}\right)^{\!2} \! \epsilon(\bu) e(\bu) = \j \frac{2\pi \c}{\lambdac} j(\bu).
\end{equation}
When $\epsilon(\bm{\cdot})$ and $\mu(\bm{\cdot})$ vary along one direction only, e.g., the $z$ direction, the coordinate system can be rotated such the variation is along $z$ and \eqref{wave_eq} becomes 
\begin{align} \notag
& \frac{1}{\mu(u_z)} \frac{\partial^2}{\partial u_x^2} e(\bu) + \frac{\partial}{\partial u_z} \left(\frac{1}{\mu(u_z)} \frac{\partial}{\partial u_z} e(\bu)\right) \\& \hspace{2cm}\label{wave_eq_z} 
+ \left(\frac{2\pi \c}{\lambdac}\right)^{\!2} \! \epsilon(u_z) e(\bu) = \j \frac{2\pi \c}{\lambdac} j(\bu).
\end{align}
An interface at $u_{0z} \ge0$ divides the medium into a lower region ${u_z > u_{0z}}$ (free-space) and an upper region ${u_z < u_{0z}}$ (material); see Fig.~\ref{fig:model2}. 
The electromagnetic properties are constant in each of the two ensuing regions, respectively specified by $(\epsilon_0,\mu_0)$ and $(\epsilon_0 \epsilon_{\sf r},\mu_0 \mu_{\sf r})$, with $\epsilon_{\sf r}$ and $\mu_{\sf r}$ as in \eqref{refractive_index}.
Following  \cite{KongBook}, we integrate \eqref{wave_eq_z} over a small rectangle centered on $\bu_0 = (u_{0x},u_{0z})$, of height $\Delta_z$ and length $\Delta_x \ll \Delta_z$, such that the field is essentially constant along $x$,
\begin{gather} \notag   
\!\!\!\!\!\!\!\!\!\!\!\!\!\!\!\!\!\!\!\!\!\!\!\!\!\!\!\!\!\!\!\!\!\!\!\! \!\!\!\!  \Delta_x \frac{1}{\mu(u_{0z}+\Delta_z)} \frac{\partial}{\partial u_z} e(u_{0x},u_{0z}+\Delta_z) \\ \notag
\!\!\!\!\!\!\!\!\!\!\!\!\!\!\!\!\!\!\!\!\!\!\!\!\!\!\!\!\!\!\!\! \!\!\!\! - \Delta_x \frac{1}{\mu(u_{0z}-\Delta_z)}  \frac{\partial}{\partial u_z} e(u_{0x},u_{0z}-\Delta_z)  \\   \label{wave_eq_z_int} 
\!\!\!\!\!\!\!\!\!\!\!\!\!\!\!\!\!\!\!\!\!\! \!\!\!\! + \left(\frac{2\pi \c}{\lambdac}\right)^{\!2} \Delta_x \int_{u_{0z}-\Delta_z}^{u_{0z}+\Delta_z} du_z \epsilon(u_z) e(u_{0x},u_z) \\   \notag 
\qquad\qquad\qquad\quad = \j \frac{2\pi \c}{\lambdac} \Delta_x \int_{u_{0z}-\Delta_z}^{u_{0z}+\Delta_z} du_z j(u_{0x},u_z).
\end{gather}
Assume the continuity of the field along $z$ such that its partial derivative is suitably defined.
From \eqref{wave_eq_z_int}, letting $\Delta_z\to0$, 
\begin{align} \label{BC_1}
\frac{1}{\mu_0} \frac{\partial}{\partial u_z} e_1(\bu_0) -  \frac{1}{\mu_0 \mu_{\sf r}} \frac{\partial}{\partial u_z} e_2(\bu_0) & = j(\bu_0) \\ \label{BC_2} 
e_2(\bu_0) & = e_1(\bu_0),
\end{align}
where $\lim_{\Delta_z\to0} \int_{u_{0z}-\Delta_z}^{u_{0z}+\Delta_z} du_z j(u_{0x},u_z) = j(\bu_0)$ for every point $\bu_0$ on the interface.
Provided there is no singular behavior across the surface, $j(\bu_0) = 0$ and \eqref{BC_1} and \eqref{BC_2} coincide with the source-free conditions in \cite[Eq.~2.1.9a]{ChewBook}.
For a dielectric material, $\mu_{\sf r}=1$, implying the continuity of the field and its normal derivative across the interface.
For a single impinging plane wave, \eqref{BC_1} and \eqref{BC_2} lead to 
 \eqref{Fresnel_spectrum} \cite{ChewBook}. 

\subsection{Physical Equivalent}

A perfectly conducting material may be regarded as an idealization where the fields inside are vanishingly small, i.e., $e_2(\bu) = 0$ \cite{KongBook,BalanisBookEM}. A current is present along a very thin layer on the surface, namely,
\begin{equation} \label{eq_surface_current}
j(\bu) = j(\bu_0) \delta(u_z-u_{0z}),
\end{equation} 
where $j(\bu_0)$ is determined by the discontinuity in the normal derivative of the field across the interface as per \eqref{BC_1},
\begin{equation} \label{BC_1_curr}
j(\bu_0) = \frac{1}{\mu_0} \frac{\partial}{\partial u_z} e_1(\bu_0) .
\end{equation}
This agrees with the uniqueness theorem applied to a volume sandwiched between the surface at $u_{0z}$ and another surface at $u_{1z}$, $u_{1z}>u_{0z}$. As $u_{1z} \to \infty$, the field over the plane at $u_{1z}$ vanishes as per radiation conditions \cite{ChewBook} and only the contribution at $u_{0z}$ matters.
In turn, the field inside such volume is expressible in terms of either the field on the surface $e_1(\bu_0)$ or its normal derivative $\frac{\partial}{\partial u_z} e_1(\bu_0)$, for every $u_{0x}$. When only the latter term is retained, the solution is said to satisfy the Neumann boundary condition \cite{ChewBook}.

Now, $e_1(\bu) = e_\sfi(\bu) + e_\sfr(\bu)$ in region~1, where $e_\sfi(\bu)$ and $e_\sfr(\bu)$ are incident and reflected fields
\cite{BalanisBookEM}.  Thus, from \eqref{BC_1_curr},
\begin{equation} \label{BC_1_curr_tot}
j(\bu_0) = \frac{1}{\mu_0} \frac{\partial}{\partial u_z} \big(e_\sfi(\bu_0) + e_\sfr(\bu_0) \big),
\end{equation}
which, for a perfect conductor, simplifies as 
\begin{equation} \label{surface_curr}
j(\bu_0) = \frac{2}{\mu_0} \frac{\partial}{\partial u_z} e_\sfi(\bu_0).
\end{equation}
For instance, consider a downgoing incident plane wave $e_\sfi(u_z) = e^{-\j \frac{2\pi}{\lambdac} u_z}$ impinging orthogonally on a far-field surface at $u_{0z}$. The upgoing reflected plane wave is given by $e_\sfr(u_z) = -e^{\j \frac{2\pi}{\lambdac} (u_z-2 u_{0z})}$. Their sum yields zero at the interface, as per $e_2(\bu) = 0$, while the sum of the plane-wave derivatives yields double each derivate's value at the interface. In the near field of the reflecting surface, the plane waves need to be replaced with Green's functions \cite[Eq.~3.6]{PlaneWaveBook}.

Plugging \eqref{surface_curr} into \eqref{eq_surface_current} yields the expression for the induced surface current density
\begin{equation} \label{surface_curr_delta}
j(\bu) = \frac{2}{\mu_0} \delta(u_z-u_{0z}) \, \frac{\partial}{\partial u_z} e_\sfi(\bu).
\end{equation}
Finally, $j(\bu)$ exists only over the portion of the interface that is actually occupied by the material,
\begin{equation} \label{surface_curr_delta_support}
j_L(\bu) = \frac{2}{\mu_0} \delta(u_z-u_{0z}) \, \I_{\left[\!-\frac{L}{2}, \frac{L}{2}\!\right]}(u_x) \, \frac{\partial}{\partial u_z} e_\sfi(\bu).
\end{equation}
A similar physical consideration is commonly used in optics to evaluate the field passing through an aperture, the so-called Kirchoff approximation \cite[Ch.~3]{OpticsBook}.
Eq. \eqref{surface_current_point} follows from \eqref{surface_curr_delta_support} by replacing $e_\sfi(\bu)$ with the Green's function $G(\bu,\bs)$ and evaluating the resulting expression at $u_{0z} = 0$ as per Fig.~\ref{fig:model2}.

\section{} \label{app:FF}

The asymptotic expansion for the Hankel function reads \cite{BalanisBookEM}
\begin{equation} \label{Hankel_approx}
H_{\nu}^{(1)}(z) \sim \sqrt{\frac{2}{\pi z}} \, e^{\j (z - \nu\frac{\pi}{2} - \frac{\pi}{4})}
\end{equation}
for every $\nu$ and large $z$ values.
Using \eqref{Hankel_approx} into \eqref{Green_fun} and \eqref{par_der_Green}, 
 \begin{align} \label{Green_asymp}
 G(\br,\bu_0) & \sim \sqrt{\frac{\j}{8 \pi}} \sqrt{\frac{\lambdac}{2\pi}} \frac{e^{\j \frac{2\pi}{\lambdac} \|\br-\bu_0\|}}{\|\br-\bu_0\|^{1/2}}  \\ \label{par_der_Green_asymp}
  \frac{\partial}{\partial u_z} G(\bu_0,\bs) & \sim \sqrt{\frac{-\j}{8 \pi}} \sqrt{\frac{2\pi}{\lambdac}} \frac{e^{\j \frac{2\pi}{\lambdac} \|\bs-\bu_0\|}}{\|\bs-\bu_0\|^{1/2}} \cos\theta_\sfi(\bu_0,\bs).
\end{align}
Now, invoke the approximations for far-zone observations \cite{BalanisBookEM}
\begin{equation} \label{}
\|\br-\bu_0\| \approx 
\begin{cases}
\|\br\| + u_x \sin\theta_\sfr(\br) & \quad \text{phase} \\
\|\br\| & \quad \text{amplitude}
\end{cases}
\end{equation}
\begin{equation} \label{}
\|\bs-\bu_0\| \approx 
\begin{cases}
\|\bs\| - u_x \sin\theta_\sfi(\bs) & \quad \text{phase} \\
\|\bs\| & \quad \text{amplitude},
\end{cases}
\end{equation}
where $\theta_\sfi(\bs) = \angle(\hat{\bz},\bs)$ and $\theta_\sfr(\br) = \angle(\hat{\bz},\br)$ are the incident angle and reflected angle corresponding to the reference point $(0,0)$ on the reflecting surface (see Fig.~\ref{fig:model2}).
These expressions can be used to convert  \eqref{Green_asymp} and \eqref{par_der_Green_asymp} into the plane waves 
 \begin{align} \label{Green_asymp_far_field}
 G(\br,\bu_0) & \sim \sqrt{\frac{\j}{8 \pi}} \sqrt{\frac{\lambdac}{2\pi}} \frac{e^{\j \frac{2\pi}{\lambdac} \|\br\|}}{\|\br\|^{1/2}}
 e^{\j \frac{2\pi}{\lambdac} u_x \sin\theta_\sfr(\br)}  \\ \label{par_der_Green_asymp_far_field}
  \frac{\partial}{\partial u_z} G(\bu_0,\bs) & \sim \sqrt{\frac{-\j}{8 \pi}} \sqrt{\frac{2\pi}{\lambdac}} \frac{e^{\j \frac{2\pi}{\lambdac} \|\bs\|}}{\|\bs\|^{1/2}} e^{-\j \frac{2\pi}{\lambdac} u_x \sin\theta_\sfi(\bs)}
  \cos\theta_\sfi(\bs).
\end{align}
Substituting \eqref{Green_asymp_far_field} and \eqref{par_der_Green_asymp_far_field} into \eqref{channel_surface} yields the far-field channel response reported in \eqref{far_field}, which follows from
$
\int_{-a/2}^{a/2} e^{-\j \beta x}  \,  dx = \left( \frac{a}{2} \right)^{\!-1} \sin(\frac{a \beta}{2}) .
$


\bibliographystyle{IEEEtran}
\bibliography{IEEEabrv,refs}

\begin{thebibliography}{10}
\providecommand{\url}[1]{#1}
\csname url@samestyle\endcsname
\providecommand{\newblock}{\relax}
\providecommand{\bibinfo}[2]{#2}
\providecommand{\BIBentrySTDinterwordspacing}{\spaceskip=0pt\relax}
\providecommand{\BIBentryALTinterwordstretchfactor}{4}
\providecommand{\BIBentryALTinterwordspacing}{\spaceskip=\fontdimen2\font plus
\BIBentryALTinterwordstretchfactor\fontdimen3\font minus
  \fontdimen4\font\relax}
\providecommand{\BIBforeignlanguage}[2]{{%
\expandafter\ifx\csname l@#1\endcsname\relax
\typeout{** WARNING: IEEEtran.bst: No hyphenation pattern has been}%
\typeout{** loaded for the language `#1'. Using the pattern for}%
\typeout{** the default language instead.}%
\else
\language=\csname l@#1\endcsname
\fi
#2}}
\providecommand{\BIBdecl}{\relax}
\BIBdecl

\bibitem{BasarSurvey}
E.~Basar~{\textit{et al.}}, ``Wireless communications through reconfigurable
  intelligent surfaces,'' \emph{IEEE Access}, vol.~7, pp. 116\,753--116\,773,
  2019.

\bibitem{ZhangSurvey}
Q.~Wu~{\textit{et al.}}, ``Intelligent reflecting surface-aided wireless
  communications: A tutorial,'' \emph{IEEE Trans. Commun.}, vol.~69, no.~5, pp.
  3313--3351, 2021.

\bibitem{Valenzuela2022}
D.~Chizhik, J.~Du, and R.~A. Valenzuela, ``Comparing power scattered by {RIS}
  with natural scatter around urban corners,'' in \emph{2022 IEEE Int. Symp.
  Antennas Propag. and USNC-URSI Radio Science Meeting (AP-S/URSI)}, 2022, pp.
  1606--1607.

\bibitem{zhang2021fundamental}
J.~Zhang, A.~A. Glazunov, W.~Yang, and J.~Zhang, ``Fundamental wireless
  performance of a building,'' \emph{IEEE Wireless Commun.}, vol.~29, no.~1,
  pp. 186--193, 2021.

\bibitem{Sciancalepore2022}
A.~Albanese~{\textit{et al.}}, ``{RIS}-aware indoor network planning: The
  {Rennes} railway station case,'' in \emph{IEEE Int. Conf. Commun.}, 2022, pp.
  2028--2034.

\bibitem{9433568}
F.~H. Danufane, M.~D. Renzo, J.~de~Rosny, and S.~Tretyakov, ``On the path-loss
  of reconfigurable intelligent surfaces: An approach based on {Green’s}
  theorem applied to vector fields,'' \emph{IEEE Trans. Commun.}, vol.~69,
  no.~8, pp. 5573--5592, 2021.

\bibitem{Emil2020}
{\"O}.~{\"O}zdogan, E.~Bj{\"o}rnson, and E.~G. Larsson, ``Intelligent
  reflecting surfaces: Physics, propagation, and pathloss modeling,''
  \emph{IEEE Wireless Commun. Lett.}, vol.~9, no.~5, pp. 581--585, 2020.

\bibitem{9941256}
Y.~Jiang, F.~Gao, M.~Jian, S.~Zhang, and W.~Zhang, ``Reconfigurable intelligent
  surface for near field communications: Beamforming and sensing,'' \emph{IEEE
  Trans. Wireless Commun.}, vol.~22, no.~5, pp. 3447--3459, 2023.

\bibitem{tulino2003capacity}
A.~M. Tulino, S.~Verdu, and A.~Lozano, ``Capacity of antenna arrays with space,
  polarization and pattern diversity,'' in \emph{IEEE Inform. Theory Workshop
  (ITW)}, 2003, pp. 324--327.

\bibitem{BalanisBookEM}
C.~A. Balanis, \emph{Advanced Engineering Electromagnetics}, 2nd~ed.\hskip 1em
  plus 0.5em minus 0.4em\relax John Wiley \& Sons, Inc., 2012.

\bibitem{Dienstfrey2001}
A.~Dienstfrey, F.~Hang, and J.~Huang, ``Lattice sums and the two-dimensional,
  periodic {G}reen's function for the {H}elmholtz equation,'' \emph{Proc. R.
  Soc. London}, vol. 457, pp. 67--85, 2001.

\bibitem{PizzoTWC22}
A.~Pizzo, A.~Lozano, S.~Rangan, and T.~L. Marzetta, ``Wide-aperture {MIMO} via
  reflection off a smooth surface,'' \emph{IEEE Trans. Wireless Commun.},
  vol.~22, no.~8, pp. 5229--5239, 2023.

\bibitem{itu_Matlab}
{International Telecommunications Union - Radiocommunications Sector (ITU-R)},
  ``{Effects of Building Materials and Structures on Radiowave Propagation
  Above 100MHz},'' ITU-R, Tech. Rep. ITU-R P.2040-1, 2015.

\bibitem{lozano2012yesterday}
A.~Lozano and N.~Jindal, ``Are yesterday's information-theoretic fading models
  and performance metrics adequate for the analysis of today's wireless
  systems?'' \emph{IEEE Commun. Mag.}, vol.~50, no.~11, pp. 210--217, 2012.

\bibitem{rappaport2010wireless}
T.~S. Rappaport, \emph{Wireless communications: Principles and practice}.\hskip
  1em plus 0.5em minus 0.4em\relax Prentice Hall, 2001.

\bibitem{Pfeiffer2013}
C.~Pfeiffer and A.~Grbic, ``Metamaterial {Huygens'} surfaces: Tailoring wave
  fronts with reflectionless sheets,'' \emph{Phys. Rev. Lett.}, vol. 110, p.
  197401, May 2013.

\bibitem{Decker2015}
M.~Decker~{\textit{et al.}}, ``High-efficiency dielectric {Huygens’}
  surfaces,'' \emph{Advanced Optical Materials}, vol.~3, no.~6, pp. 813--820,
  2015.

\bibitem{ChewBook}
W.~C. Chew, \emph{Waves and Fields in Inhomogenous Media}.\hskip 1em plus 0.5em
  minus 0.4em\relax Wiley-IEEE Press, 1995.

\bibitem{HeedongIRS}
H.~Do, N.~Lee, and A.~Lozano, ``Line-of-sight {MIMO} via intelligent reflecting
  surface,'' \emph{IEEE Trans. Wireless Commun.}, vol.~22, no.~6, pp.
  4215--4231, 2023.

\bibitem{PlaneWaveBook}
T.~B. Hansen and A.~D. Yaghjian, \emph{Plane-Wave Theory of Time-Domain
  Fields}.\hskip 1em plus 0.5em minus 0.4em\relax Wiley-IEEE Press, 1999.

\bibitem{degli2022reradiation}
V.~Degli-Esposti, E.~M. Vitucci, M.~Di~Renzo, and S.~A. Tretyakov,
  ``Reradiation and scattering from a reconfigurable intelligent surface: A
  general macroscopic model,'' \emph{IEEE Trans. Antennas and Propag.},
  vol.~70, no.~10, pp. 8691--8706, 2022.

\bibitem{KongBook}
J.~A. Kong, \emph{Electromagnetic Wave Theory}.\hskip 1em plus 0.5em minus
  0.4em\relax EMW Publishing, 2008.

\bibitem{OpticsBook}
M.~Born and E.~Wolf, Eds., \emph{Principles of Optics}, sixth edition~ed.\hskip
  1em plus 0.5em minus 0.4em\relax Pergamon Press, 1980.

\end{thebibliography}

\end{document}